\documentclass[epj]{svjour}
\usepackage{amsfonts,amssymb,amsmath}
\usepackage{graphicx,color}
\usepackage{dcolumn,slashed}
\usepackage{adjustbox}
\usepackage{braket}

\makeatletter
\DeclareRobustCommand{\cev}[1]{
  {\mathpalette\do@cev{#1}}
}
\newcommand{\do@cev}[2]{
  \vbox{\offinterlineskip
    \sbox\z@{$\m@th#1 x$}
    \ialign{##\cr
      \hidewidth\reflectbox{$\m@th#1\vec{}\mkern4mu$}\hidewidth\cr
      \noalign{\kern-\ht\z@}
      $\m@th#1#2$\cr
    }
  }
}
\makeatother

\begin{document}

\title{Impurity Lattice Monte Carlo for Hypernuclei}

{\color{red}
\author{Dillon Frame \inst{1} 
\and Timo A. L\"{a}hde \inst{1} 
\and Dean Lee\inst{2}  
\and Ulf-G.~Mei{\ss}ner\inst{3,1,4}
}                     
\institute{
Institut~f\"{u}r~Kernphysik,~Institute~for~Advanced~Simulation and
J\"{u}lich~Center~for~Hadron~Physics, \\ 
Forschungszentrum~J\"{u}lich, D-52425~J\"{u}lich,~Germany
\and Facility for Rare Isotope Beams and Department of Physics and Astronomy, Michigan State University, \\ 
MI 48824, USA
\and Helmholtz-Institut~f\"{u}r~Strahlen-~und~Kernphysik~and~Bethe~Center~for
Theoretical~Physics, Universit\"{a}t~Bonn, \\ D-53115~Bonn,~Germany
\and Tbilisi State University, 0186 Tbilisi, Georgia
}
}

\date{Received: date / Revised version: date}
%

\abstract{
We consider the problem of including $\Lambda$ hyperons into the {\it ab initio} framework of nuclear
lattice effective field theory. In order to avoid large sign oscillations in Monte Carlo simulations,
we make use of the fact that the number of hyperons is typically small compared to the number of nucleons in
the hypernuclei of interest. This allows us to use the impurity lattice Monte Carlo method, where the minority species
of fermions in the full nuclear Hamiltonian is integrated out and  treated as a worldline in Euclidean projection time.
The majority fermions (nucleons) are treated as explicit degrees of freedom, with
their mutual interactions described by auxiliary fields.  This is the first application of the impurity lattice Monte Carlo method to systems where the majority particles are interacting.  Here, we show how the impurity Monte Carlo method can be applied to compute the binding energy of the light hypernuclei.  In this exploratory work we use spin-independent nucleon-nucleon and hyperon-nucleon interactions to test the computational power of the method.  We find that the computational effort scales approximately linearly in the number of nucleons.  The results are very promising for future studies of larger hypernuclear systems using chiral effective field theory and realistic hyperon-nucleon interactions, as well as applications to other quantum many-body systems. 
} 

\maketitle


\section{Introduction}

Hypernuclei are bound states of one or two hyperons together with a core composed of nucleons. They extend the nuclear chart into a third dimension, augmenting the usual two dimensions of proton number and neutron number. We will use the notation $Y$ for a $\Lambda$ or $\Sigma$ hyperon and $N$ for a nucleon.
Due to the scarcity of direct hyperon-nucleon ($YN$) and hyperon-hyperon ($YY$) scattering data, these unusual forms of baryonic matter play
an important role in
pinning down the fundamental baryon-baryon forces. This requires on the one hand an effective field theory (EFT)
description of the underlying forces, as pioneered in Ref.~\cite{Korpa:2001au,Polinder:2006zh}, and on the other hand a numerically precise and consistent method to solve the nuclear $A$-body problem, such as nuclear lattice EFT (NLEFT)
\cite{Lee:2008fa,Lahde:2019npb}. For calculations combining these chiral EFT forces at LO and NLO
\cite{Haidenbauer:2013oca,Haidenbauer:2019boi} with other
many-body methods, see \textit{e.g}. Ref.~\cite{Lonardoni:2013rm,Gazda:2016qva,Wirth:2017lso,Wirth:2017bpw,Le:2019gjp,Haidenbauer:2019thx}.

In view of the success of NLEFT in the description of nuclear spectra and reactions, it seems natural
to extend this method to hypernuclei. However, this is not quite straightforward. While one can extend the four spin-isospin degrees of freedom comprising the nucleons to include the $\Lambda$ and $\Sigma$ states~\cite{Bour},
this has not been done because there is no longer an approximate symmetry such as Wigner's SU(4) symmetry \cite{Wigner:1936dx} that protects the
Monte Carlo (MC) simulations against strong sign oscillations when using auxiliary fields.\footnote{In the SU(3) limit of equal up, down and strange quark masses, such a spin-flavor symmetry might be restored \cite{Wagman:2017tmp}, but this limit is far from the physical world.} The physics of hypernuclei therefore requires a different approach, and in this paper we show how the computational problems are solved 
using the impurity lattice Monte Carlo (ILMC) method. 

The ILMC method was introduced in Ref.~\cite{Elhatisari:2014lka} in the context of a Hamiltonian 
theory of spin-up and spin-down fermions, and applied to the intrinsically non-perturbative physics of
Fermi polarons in two dimensions in Ref.~\cite{Bour:2014bxa}.
The ILMC method is particularly useful for the case where only one fermion (of either species) is immersed
in a ``sea'' of the other species. Within the standard auxiliary
field Monte Carlo method, such an extreme imbalance would lead to unacceptable sign oscillations
in the Monte Carlo probability weight. In the ILMC method, the minority particle is ``integrated out'',
resulting in a formalism where only the majority species fermions appear as explicit degrees of freedom, while
the minority fermion is represented by a ``worldline'' in Euclidean projection time. The spatial position of
this worldline is updated using Monte Carlo updates, while the interactions between the majority fermions are
described by the auxiliary field formalism~\cite{Lahde:2019npb}.

Here, we apply the ILMC method to the inclusion of hyperons into NLEFT
simulations. We identify the $\Lambda$ hyperon as the minority species, which we represent by a worldline
in Euclidean time. This
$\Lambda$ worldline is treated as immersed in an environment consisting of some number of nucleons.
We focus on the Monte Carlo calculation of the binding energy of light hypernuclei, by means of a simplified
$YN$ interaction, consisting of a single contact interaction, tuned to a best description of the the empirical binding
energies of the $s$-shell hypernuclei with $A=3,4,5$.\footnote{We are well aware of the importance of
the  $\Lambda N$-$\Sigma N$ transition.
However, we choose a simple starting point for this exploratory study and will consider more realistic interactions in a later publication.}
For the $NN$ interaction, we use a simple leading order interaction similar to that described in Ref.~\cite{Lu:2018bat}. We benchmark our ILMC
results against Lanczos calculations of transfer matrix and 
exact Euclidean projection calculations with initial/final states and number of time steps that match the ILMC calculations.
We note that our Monte Carlo method is free from any approximation about the nodal structure of the many-body wave function. This is the first application of such unconstrained Monte Carlo simulations to hypernuclei.

This paper is organized as follows. In Sec.~\ref{sec:form}, we present the path integral formalism for our system of nucleons and one hyperon.  We first write the nucleon-nucleon interaction first without auxiliary fields and then with auxiliary fields.  In Sec.~\ref{sec:transfer} we present the equivalent system using normal-ordered transfer matrices.  In Sec.~\ref{sec:imp}, we derive the impurity worldline formalism for the chosen $YN$ interaction, and introduce the concept of
the ``reduced'' transfer matrix operator, which acts on
the nucleons only. In Sec.~\ref{sec:mc}, we discuss the Monte Carlo updating of the
hyperon worldline and the auxiliary fields, which encode the interactions between nucleons.
In Sec.~\ref{sec:res}, we present results for the ground state energies of the $s$-shell nuclei and hypernuclei. 
In Sec.~\ref{sec:disc}, we conclude with a discussion of future improvements and applications of the impurity lattice Monte Carlo method to hypernuclei and other quantum many-body systems.


\section{Path integral formalism \label{sec:form}}

We develop the ILMC formalism following Ref.~\cite{Elhatisari:2014lka}, who considered
a system of spin-up and spin-down fermions, with a contact interaction which operates between
fermions of opposite spin. The situation here is completely analogous, we have one majority
species, the nucleons, and {\em one} impurity, the $\Lambda$.
As usual in NLEFT, we consider positions on a spatial lattice denoted by $\vec n$ and
lattice spacing $a$. We also assume that Euclidean time has been discretized, such that slices of
the Euclidean time are denoted by $n_t$ with temporal lattice spacing
$a_t$. The partition function can be expressed in terms of the Grassmann path integral
\begin{align}
\mathcal{Z} = \int 
\Bigg[\prod_{\substack{\vec n, n_t \\ s=N,Y}}
d\zeta_s^{}(\vec n, n_t^{}) d\zeta_s^*(\vec n, n_t^{}) \Bigg]
\exp(-S[\zeta, \zeta^*]),
\label{Z_pi}
\end{align}
where the subscripts $N$ refer to all nucleon spin and isospin components and $Y$ refers to all hyperon spin components.  
In this study we consider only $\Lambda$ hyperons.  In future work we will also consider $\Sigma$ hyperons 
or account for their influence via three-baryon interactions involving a $\Lambda$ and two nucleons.  
We also make the simplifying assumption that the hyperon-nucleon and nucleon-nucleon interaction 
are spin-independent and neglect Coulomb interactions.  Because of the spin-independent interaction and the fact that we have only one Lambda hyperon, from this point onward we can restrict our attention to only one spin component of the hyperon.

Assuming that the  exponent of the Euclidean action in Eq.~(\ref{Z_pi}) is treated by a
Trotter decomposition, we find
\begin{align}
& S[\zeta, \zeta^*] \equiv \sum_{n_t} \bigg\{
S_t^{}[\zeta, \zeta^*,n_t^{}] +
S_Y^{}[\zeta^{}, \zeta^*,n_t^{}] 
\nonumber \\
& \quad + S_N^{}[\zeta^{}, \zeta^*,n_t^{}] 
+ S_{YN}^{}[\zeta, \zeta^*,n_t^{}] + S_{NN}^{}[\zeta, \zeta^*,n_t^{}]
\bigg\},
\end{align}
where the component due to the time derivative is
\begin{align}
S_t^{}[\zeta, \zeta^*,n_t^{}] & \equiv 
\!\!\! \sum_{\vec n,s = N,Y} 
\zeta_s^*(\vec n,n_t^{})
\nonumber \\
& \quad \times \bigg[ \zeta_s^{}(\vec n,n_t^{}+1) - \zeta_s^{}(\vec n,n_t^{}) \bigg],
\end{align}
while $S_Y^{}$ and $S_N^{}$ describe the kinetic energies of the hyperons and nucleons, respectively.
Further, $S_{YN}$ provides the $YN$ interaction, and $S_{NN}$
the $NN$ interaction, which we shall consider next.


\subsection{The hyperon-nucleon interaction}

For the hyperons, we take for simplicity the lowest-order (unimproved) kinetic energy
\begin{align}
& S_Y^{}[\zeta, \zeta^*,n_t^{}] \equiv 6h \sum_{\vec n} 
\zeta_Y^*(\vec n,n_t^{}) \zeta_Y^{}(\vec n,n_t^{})
\nonumber \\
& - h \sum_{\vec n} 
\sum_{l = 1}^3 \: \zeta_Y^*(\vec n,n_t^{})
\bigg[ \zeta_Y^{}(\vec n+\hat e_l^{},n_t^{}) + \zeta_Y^{}(\vec n-\hat e_l^{},n_t^{}) \bigg],
\label{hyp_kin}
\end{align}
with
\begin{equation}
h \equiv \frac{\alpha_t^{}}{2m_Y^{}},
\end{equation}
where $m_Y$ is the hyperon mass, and
we have defined $\alpha_t^{} \equiv a_t^{}/a$ as the ratio of temporal and spatial lattice spacings.

The $YN$ interaction is given by
\begin{align}
& S_{YN}^{}[\zeta, \zeta^*,n_t^{}] \equiv
\alpha_t^{} C_{YN}^{} \sum_{\vec n}
\rho_N^{}(\vec n, n_t^{}) \rho_Y^{}(\vec n, n_t^{}),
\label{YN_Grassmann}
\end{align}
where
\begin{align}
\rho_N^{}(\vec n, n_t^{}) 
\equiv \sum_{i,j} \rho_{i,j}^{}(\vec n, n_t^{})
\equiv \sum_{i,j} \zeta^*_{i,j}(\vec n, n_t^{}) \zeta^{}_{i,j}(\vec n, n_t^{}),
\label{rho_nucl}
\end{align}
and
\begin{equation}
\rho_Y^{}(\vec n, n_t^{}) \equiv \zeta_Y^*(\vec n, n_t^{}) \zeta_Y^{}(\vec n, n_t^{}),
\label{rho_hyp}
\end{equation}
are nucleon and hyperon densities, respectively, with spin $i=0,1$
(up, down) and isospin $j=0,1$ (proton, neutron). 
The tuning of the coupling constant $C_{YN}$ is discussed in Section~\ref{sec:res}.

Note that this is a simplified version of the pionless EFT calculation of Ref.~\cite{Hammer:2001ng}, 
which also included a three-body interaction at LO. Such an interaction is sub-leading in
chiral EFT approaches (such as NLEFT). See also the recent work in Ref.~\cite{Contessi:2018qnz}.


\subsection{The nucleon-nucleon interaction}

For the kinetic energy of the nucleon degrees of freedom, we likewise use the lowest-order expression
\begin{align}
& S_N^{}[\zeta, \zeta^*,n_t^{}] \equiv
\frac{3\alpha_t^{}}{m_N^{}}\sum_{\vec n} \rho_N^{}(\vec n, n_t^{})
\nonumber \\
& -\frac{\alpha_t^{}}{2m_N^{}}
\sum_{\vec n} \sum_{l = 1}^3 
\left[\rho_N^{}(\vec n, \vec n+\hat e_l^{}, n_t^{}) + \rho_N^{}(\vec n, \vec n-\hat e_l^{}, n_t^{})\right],
\label{kin_nucl}
\end{align}
where 
\begin{align}
\rho_N^{}(\vec n, \vec n^\prime, n_t^{}) & \equiv \sum_{i,j} 
\zeta^*_{i,j}(\vec n, n_t^{}) \zeta^{}_{i,j}(\vec n^\prime, n_t^{}), \\
\rho_N^{}(\vec n, n_t^{}) & \equiv \rho_N^{}(\vec n, \vec n, n_t^{}),
\end{align}
and $m_N^{}$ is the nucleon mass. Here, the $\hat e_l^{}$ are unit vectors
in lattice direction $l$.

The Wigner SU(4)-symmetric part of
the leading-order (LO) $NN$ interaction of Refs.~\cite{Elhatisari:2016owd,Elhatisari:2017eno,Li:2018ymw} 
is used for the present work. This is an approximate
symmetry~\cite{Wigner:1936dx} of the low-energy nucleon-nucleon interactions,
where the spin and isospin degrees of freedom of the nucleons can be rotated as
four components of an SU(4) multiplet. We have
\begin{align}
S_{NN}^{}[\zeta, \zeta^*,n_t^{}] & \equiv \frac{\alpha_t^{} C_{NN}^{}}{2} \! \! \! \sum_{\vec n, \vec n^\prime, \vec n^{\prime\prime}}
\rho_N^s(\vec n^\prime,n_t^{}) f_{s_{\rm L}^{}}^{}(\vec n^\prime - \vec n)
\nonumber \\
& \quad \times 
f_{s_{\rm L}^{}}^{}(\vec n - \vec n^{\prime\prime}) \rho_N^s(\vec n^{\prime\prime},n_t^{}),
\label{SNN}
\end{align}
where
\begin{align}
\rho_N^s(\vec n, n_t^{}) \equiv 
\sum_{i,j} 
\zeta^{s_{\rm NL}^{}*}_{i,j}(\vec n, n_t^{})
\zeta^{s_{\rm NL}^{}}_{i,j}(\vec n, n_t^{}),
\label{rho_nucl_smear}
\end{align}
is the smeared nucleon density, and the (local) smearing function
$f_{s_{\rm L}}$ is defined as
\begin{align}
f_{s_{\rm L}^{}}(\vec n) & \equiv 1 \; {\rm for} \; | \vec n | = 0, \nonumber \\
& \equiv s_{\rm L}^{} \; {\rm for} \; | \vec n | = 1, \nonumber \\
& \equiv 0 \; {\rm otherwise},
\end{align} 
and the (non-locally) smeared Grassmann fields are given by
\begin{equation}\label{smear1}
\zeta^{s_{\rm NL}^{}}_{i,j}(\vec n, n_t^{}) \equiv
\zeta_{i,j}^{}(\vec n, n_t^{}) + s_{\rm NL}^{} \sum_{|\vec n^\prime | = 1} 
\zeta_{i,j}^{}(\vec n + \vec n^\prime, n_t^{}),
\end{equation}
and
\begin{equation}\label{smear2}
\zeta^{s_{\rm NL}^{}*}_{i,j}(\vec n, n_t^{}) \equiv 
\zeta^*_{i,j}(\vec n, n_t^{}) + s_{\rm NL}^{} \sum_{|\vec n^\prime | = 1}
\zeta^*_{i,j}(\vec n + \vec n^\prime, n_t^{}),
\end{equation}
where the values of the parameters $C_{NN}$, $s_{\rm L}$ and $s_{\rm NL}$ used for the present work
are discussed in Section~\ref{sec:res} (see also Ref.~\cite{Lu:2018bat} for a full treatment). 

For the $NN$ interaction we can reduce the expressions quadratic in the nucleon densities using
the relation
%
\begin{align}
& \exp\left(-\frac{\alpha_t^{} C_{NN}^{}}{2} \tilde\rho^2
\right) = & \: \nonumber \\
& \frac{1}{\sqrt{2\pi}}\int_{-\infty}^\infty d\phi \, e^{-\frac{\phi^2}{2}}
 \exp\left(\sqrt{-\alpha_t^{} C_{NN}^{}} \, 
\phi \tilde\rho \right),
\end{align}
where
\begin{equation}
\tilde\rho \equiv \sum_{\vec n^\prime}
f_{s_{\rm L}^{}}^{}(\vec n - \vec n^{\prime}) \rho_N^s(\vec n^{\prime}, n_t^{}),
\end{equation}
for each lattice site ($\vec n, n_t^{}$), 
such that $\phi(\vec n, n_t^{})$ is treated as a scalar auxiliary (Hubbard-Stratonovich) field.
The $NN$ action then becomes
\begin{align}
& \exp(-S_{NN}^{}[\zeta, \zeta^*,n_t^{}]) \nonumber \\
& = \int \prod_{\vec n} \left[\frac{d\phi(\vec n,n_t^{})}{\sqrt{2\pi}} 
e^{-\frac{1}{2} \phi^2(\vec n,n_t^{})}\right]\exp(-S_{\phi N}^{}[\zeta, \zeta^*,n_t^{}]),
\label{HS1}
\end{align}
for Euclidean time slice $n_t^{}$,
where
\begin{align}
& S_{\phi N}^{}[\zeta, \zeta^*,n_t^{}] =
-\sqrt{-\alpha_t^{} C_{NN}^{}} \nonumber \\ 
& \qquad \times \sum_{\vec n, \vec n^\prime} \phi(\vec n, n_t^{})
f_{s_{\rm L}^{}}^{}(\vec n - \vec n^\prime)
\rho_N^s(\vec n^\prime, n_t^{}),
\label{HS2}
\end{align}
for $C_{NN} < 0$.

In the ILMC calculations, the path integral over the auxiliary field
$\phi$ is evaluated using either local Metropolis algorithm updates or global lattice updates using the hybrid Monte Carlo (HMC) algorithm. 
See Ref.~\cite{Lu:2018bat} for more details on efficient Monte Carlo algorithms.

\section{Transfer matrix formalism \label{sec:transfer}}

Derivations of Feynman rules are usually easier to perform in the Grassmann 
formalism. However, actual NLEFT calculations are performed using the transfer matrix Monte Carlo method.
As noted in Ref.~\cite{Elhatisari:2014lka}, the Grassmann and transfer matrix operator 
formulations are connected by the exact relationship
\begin{align}
& \mathrm{Tr} \big\{ : f_{N_t-1}^{}[a_s^{}(\vec n),a_{s^\prime}^\dagger(\vec n^\prime)] : 
\cdots : f_0^{}[a_s^{}(\vec n),a_{s^\prime}^\dagger(\vec n^\prime)] : \big\} =
\nonumber \\
& \int \Bigg[
\prod_{\substack{\vec n, n_t \\ s=N,Y}}
d\zeta_s^{}(\vec n, n_t^{}) d\zeta_s^*(\vec n, n_t^{}) \Bigg]
\exp\left(-\sum_{n_t} S_t^{}[\zeta, \zeta^*, n_t^{}]\right)
\nonumber \\
& \quad \times \prod_{n_t=0}^{N_t-1}
f_{n_t}^{}\big[\zeta_s^{}(\vec n, n_t^{}), \zeta_{s^\prime}^*(\vec n^\prime, n_t^{})\big],
\label{relation_3}
\end{align}
where $f$ is an arbitrary function, $a_s^\dagger$ and $a_s^{}$ denote creation and annihilation operators for
the fermion degrees of freedom, and the colons
signify normal ordering.  Using this identity, we can write the partition function in Eq.~(\ref{Z_pi}) as 
\begin{equation}
\mathcal{Z} = \mathrm{Tr}(\hat M^{N_t^{}}),
\label{Z_m}
\end{equation}
where $\hat M$ is the (normal-ordered) transfer matrix operator.

We can use Eq.~(\ref{relation_3}) to define the full transfer matrix operator as
\begin{equation}
\hat M = \, : \exp(-\alpha_t^{}\hat H) :.
\label{ham1}
\end{equation}
with Hamiltonian
\begin{equation}
\hat H \equiv \hat H_0^N + \hat H_0^Y + \hat H_{NN}^{} + \hat H_{YN}^{}.
\label{ham2}
\end{equation}
We now go through each of these terms. The nucleon kinetic energy term is
\begin{align}
\hat H_0^N & \equiv
\frac{3}{m_N^{}}\sum_{\vec n} \hat\rho_N^{}(\vec n)
\nonumber \\
& -\frac{1}{2m_N^{}}
\sum_{\vec n} \sum_{l = 1}^3 
\left[\hat\rho_N^{}(\vec n, \vec n+\hat e_l^{}) + \hat\rho_N^{}(\vec n, \vec n-\hat e_l^{})\right],
\label{kin_nucl_op}
\end{align}
with 
\begin{align}
\hat\rho_N^{}(\vec n, \vec n^\prime) & \equiv \sum_{i,j} 
a^{\dagger}_{i,j}(\vec n) a^{}_{i,j}(\vec n^\prime), \\
\hat\rho_N^{}(\vec n) & \equiv \hat\rho_N^{}(\vec n,\vec n).
\end{align}
The hyperon kinetic energy term is 
\begin{align}
\hat H_0^N & \equiv
\frac{3}{m_Y^{}}\sum_{\vec n} \hat\rho_Y^{}(\vec n)
\nonumber \\
& -\frac{1}{2m_Y^{}}
\sum_{\vec n} \sum_{l = 1}^3 
\left[\hat\rho_Y^{}(\vec n, \vec n+\hat e_l^{}) + \hat\rho_Y^{}(\vec n, \vec n-\hat e_l^{})\right],
\end{align}
with 
\begin{align}
\hat\rho_Y^{}(\vec n, \vec n^\prime) & \equiv \sum_{i,j} 
a^{\dagger}_Y(\vec n) a^{}_Y(\vec n^\prime), \\
\hat\rho_Y^{}(\vec n) & \equiv \hat\rho_Y^{}(\vec n,\vec n).
\end{align}
The $NN$ interaction is
\begin{align}
\hat H_{NN}^{} & = \frac{C_{NN}^{}}{2}: \! \! \! \sum_{\vec n, \vec n^\prime, \vec n^{\prime\prime}}
\hat\rho_N^s(\vec n^\prime) f_{s_{\rm L}^{}}^{}(\vec n^\prime - \vec n)
\nonumber \\
& \quad \times 
f_{s_{\rm L}^{}}^{}(\vec n - \vec n^{\prime\prime}) \hat\rho_N^s(\vec n^{\prime\prime}):,
\label{VNN}
\end{align}
with
\begin{align}
\hat\rho_N^s(\vec n) \equiv \sum_{i,j} 
a^{s_{\rm NL}^{}\dagger}_{i,j}(\vec n)
a^{s_{\rm NL}^{}}_{i,j}(\vec n),
\label{rho_nucl_smear_op}
\end{align}
and the operators 
$a^{s_{\rm NL}^{}\dagger}_{i,j}(\vec n)$ and
$a^{s_{\rm NL}^{}}_{i,j}(\vec n)$
are defined in terms of the (non-locally) smeared annihilation and creation operators
\begin{equation}\label{smear1_op}
a^{s_{\rm NL}^{}}_{i,j}(\vec n) \equiv
a_{i,j}^{}(\vec n) + s_{\rm NL}^{} \sum_{|\vec n^\prime | = 1} 
a_{i,j}^{}(\vec n + \vec n^\prime),
\end{equation}
and
\begin{equation}\label{smear2_op}
a^{s_{\rm NL}^{}\dagger}_{i,j}(\vec n) \equiv 
a^{\dagger}_{i,j}(\vec n) + s_{\rm NL}^{} \sum_{|\vec n^\prime | = 1}
a^{\dagger}_{i,j}(\vec n + \vec n^\prime).
\end{equation}
The $YN$ interaction is
\begin{equation}
\hat H_{YN}^{} = C_{YN}^{} \sum_{\vec n} \hat \rho_N^{}(\vec n) \hat \rho_Y^{}(\vec n).
\end{equation}

When rewriting the nucleon-nulceon interaction with auxiliary fields, the partition function takes the form
\begin{align}
\mathcal{Z} = 
\int &\prod_{\vec n,n_t} \left[\frac{d\phi(\vec n,n_t^{})}{\sqrt{2\pi}} 
e^{-\frac{1}{2} \phi^2(\vec n,n_t^{})}\right]
\mathrm{Tr}[\hat M^{(N_t-1)} \cdots \hat M^{(0)}],
\end{align}
where
\begin{equation}
\hat M^{(n_t)} \equiv \, : \exp(-\alpha_t^{}\hat H^{(n_t)}) :,
\end{equation}
with
\begin{equation}
\hat H^{(n_t)} \equiv \hat H_0^N + \hat H_0^Y + \hat H^{(n_t)}_{\phi N} + \hat H_{YN}^{},
\end{equation}
and
\begin{align}
& \hat H^{(n_t)}_{\phi N}=
-\sqrt{-\alpha_t^{} C_{NN}^{}} \nonumber \\ 
& \qquad \times \sum_{\vec n, \vec n^\prime} \phi(\vec n, n_t^{})
f_{s_{\rm L}^{}}^{}(\vec n - \vec n^\prime)
\hat \rho_N^s(\vec n^\prime, n_t^{}).
\end{align}

\section{Impurity worldlines and reduced transfer matrices \label{sec:imp}}

We now integrate out the hyperon degree of freedom and
derive a ``reduced'' transfer matrix operator $\hat{\slashed M}$, which acts on the nucleon degrees of freedom only.  Let us consider the transfer matrix between time slices $n_t$ and $n_t+1$.  Let $\ket{\vec{n}}$ represent the state with the hyperon at lattice site $\vec{n}$.  We first consider the case when the hyperon hops from lattice site $\vec{n}$ to $\vec{n}\pm \hat e_l$.  We then have
\begin{align}
    \braket{\vec{n}\pm \hat e_l | \hat M^{(n_t)} | \vec{n}} = {\hat {\slashed M}}^{(n_t)}_{\vec{n}\pm \hat e_l,\vec{n}}
\end{align}
where ${\hat {\slashed M}}^{(n_t)}_{\vec{n}\pm \hat e_l,\vec{n}}$ is the reduced transfer matrix operator acting on only the nucleons with
\begin{equation}
    \hat{\slashed M}^{(n_t)}_{\vec{n}\pm \hat e_l,\vec{n}}
    = h:\exp(-\alpha_t{\hat H}^{(n_t)}_{\vec{n}\pm \hat e_l,\vec{n}}):, \label{Mslash_hop}
\end{equation}
where
\begin{equation}
    {\hat H}^{(n_t)}_{\vec{n}\pm \hat e_l,\vec{n}} = H_0^N + \hat H_0^Y + \hat H^{(n_t)}_{\phi N}.
\end{equation}

Next we consider the case when the hyperon remains at lattice site $\vec{n}$ between time slices $n_t$ and $n_t+1$.  We then have
\begin{align}
    \braket{\vec{n} | \hat M^{(n_t)} | \vec{n}} = {\hat {\slashed M}}^{(n_t)}_{\vec{n},\vec{n}},
\end{align}
where the reduced transfer matrix is
\begin{equation}
    \hat{\slashed M}^{(n_t)}_{\vec{n},\vec{n}}
    = (1-6h):\exp(-\alpha_t{\hat H}^{(n_t)}_{\vec{n},\vec{n}}):, \label{Mslash_stat}
\end{equation}
with
\begin{equation}
    {\hat H}^{(n_t)}_{\vec{n},\vec{n}} = H_0^N + \hat H_0^Y + \hat H^{(n_t)}_{\phi N}+\frac{C_{YN}^{}}{1-6h} \hat\rho_N^{}(\vec n) + \cdots.
\end{equation}
The ellipses refers to terms with higher powers of $\hat \rho$ and additional factors of $\alpha_t$.  These are lattice artifacts that disappear in the limit $\alpha_t \rightarrow 0$.  They are needed to cancel the higher-order powers of the $C_{YN}$ term when expanding the exponential in Eq.~(\ref{Mslash_stat}) beyond the linear term.  In the full transfer matrix such terms vanish upon normal ordering of the hyperon field since we have only one hyperon in our system.  However, when we integrate out the hyperon worldline, such terms no longer vanish since the hyperon is no longer a dynamical field.   

In our simulations here we drop all such higher-order terms from our ILMC simulations.  This choice constitutes a redefinition of our starting interaction to include some small higher-body interactions between the hyperon and more than one nucleon.  Since we will take $\alpha_t$ to be very small, the most important induced higher-body interaction is a small three-body interaction.  The three-body interaction has the form
\begin{equation}
\hat H_{YNN}^{} = - \frac{\alpha_t^{} C_{YN}^2}{2(1-6h)} \sum_{\vec n} 
\hat \rho_N^{}(\vec n) \hat \rho_N^{}(\vec n) \hat \rho_Y^{}(\vec n),
\label{H_spurious}
\end{equation}
We see explicitly that this term is a lattice artifact that disappears when $\alpha_t^{} \to 0$. 


\section{Monte Carlo calculation \label{sec:mc}}

We now describe how ILMC calculations are performed using the Projection Monte Carlo (PMC) method.
Let us first assume that the impurity has been fixed at a given spatial lattice site, and that no
``hopping'' of the impurity occurs during the Euclidean time evolution. We shall then relax this constraint, and
discuss a practical algorithm for updating the configuration of the hyperon worldline.

\subsection{Stationary impurity \label{sec:mc_stat}}

For a stationary hyperon impurity, the
reduced transfer matrix is given by Eq.~(\ref{Mslash_stat}), and for the purposes of the PMC calculation, we
define the Euclidean projection amplitude
\begin{equation}
Z_{jk}^{}(N_t^{}) \equiv \langle \psi_j^{} | \hat{\slashed M}^{N_t^{}} | \psi_k^{} \rangle,
\end{equation}
for a product of $N_t^{}$ Euclidean time slices,
where $j$ and $k$ denote different initial cluster states.
As usual, this is expressed as a determinant of single-particle amplitudes, which gives
\begin{equation}
Z_{jk}^{}(N_t^{}) = \mathrm{det} \, M_{p\times p}^{jk},
\label{proj_ampl}
\end{equation}
where
\begin{equation}
M_{p\times p}^{jk} = \left(
\begin{array}{c c c}
\vspace{.1cm}
\langle \phi_{0,j}^{} | \hat{\slashed M}^{N_t^{}} | \phi_{0,k}^{} \rangle & 
\langle \phi_{0,j}^{} | \hat{\slashed M}^{N_t^{}} | \phi_{1,k}^{} \rangle & \cdots \\
\vspace{.1cm}
\langle \phi_{1,j}^{} | \hat{\slashed M}^{N_t^{}} | \phi_{0,k}^{} \rangle &
\langle \phi_{1,j}^{} | \hat{\slashed M}^{N_t^{}} | \phi_{1,k}^{} \rangle & \cdots \\
\vdots & \vdots & \ddots
\end{array}
\right),
\end{equation}
for $p$ nucleons. By means of the projection amplitudes~(\ref{proj_ampl}), we construct
\begin{equation}
[\hat M^a_{}(N_t^{})]_{qq^\prime}^{} \equiv 
\sum_{q^{\prime\prime}}Z_{qq^{\prime\prime}}^{-1}(N_t^{})
Z_{q^{\prime\prime}q^\prime}^{}(N_t^{}+1),
\label{M_adiab}
\end{equation}
which is known as the ``adiabatic transfer matrix''. If we denote the eigenvalues of~(\ref{M_adiab}) by 
$\lambda_i(N_t)$, we find
\begin{equation}
\lambda_i^{}(N_t^{}) = \exp(-\alpha_t^{}E_i^{}(N_t^{}+1/2)),
\end{equation}
such that the low-energy spectrum is given by the ``transient'' energies
\begin{equation}
E_i^{}(N_t^{}+1/2) = -\frac{\log(\lambda_i^{}(N_t^{}))}{\alpha_t^{}},
\end{equation}
at finite temporal lattice spacing $a_t$. For the case of a single trial cluster state with $p$ nucleons, 
Eq.~(\ref{proj_ampl}) reduces to 
\begin{equation}
Z(N_t^{}) = \mathrm{det} \, M_{p\times p}^{00},
\end{equation}
for the case of a single trial state. 
The ground-state energy is obtained from
\begin{equation}
E_0^{}(N_t^{}+1/2) = -\frac{\log(Z(N_t^{}+1)/Z(N_t^{}))}{\alpha_t^{}},
\end{equation}
in the limit $N_t \to \infty$, where
the exact low-energy spectrum of the transfer matrix will be recovered.
Note that the argument $N_t+1/2$ is conventionally
assigned to the transient energy computed from the ratio of projection amplitudes evaluated at
Euclidean time steps $N_t+1$ and $N_t$. 

As an example, for the hypertriton we have $p = 2$ nucleons after the impurity hyperon has been integrated out.
We start the Euclidean time projection with a single initial trial cluster state ($j = k = 0$)
consisting of a spin-up proton and a spin-up neutron. 
As there are no terms that mix spin or isospin, the other components of each single-particle state are set to zero, and 
remain so during the PMC calculation. For the spatial parts of the nucleon wave functions, we may choose, for example, the zero-momentum state
\begin{equation}
| \phi_{0,0}^{} \rangle = | \phi_{1,0}^{} \rangle = \langle 0,0,0 \rangle, 
\end{equation}
in the notation of Ref.~\cite{Elhatisari:2014lka}, which denotes 
plane-wave orbitals in a cubic box. In principle, we may also choose any other plane-wave state with 
non-zero momentum (see Table~1 of Ref.~\cite{Elhatisari:2014lka}), or any other more complicated trial state.  For the heavier nuclei, it is indeed better to choose an initial state where the nucleons are clustered together.  In this case we sum over all possible translations of the cluster in order construct an initial state with zero total momentum.

\subsection{Hopping impurity \label{sec:mc_hop}}

If the hyperon impurity is allowed to hop between nearest-neighbor sites (from one Euclidean time slice to the next),
the Euclidean projection amplitude becomes a sum over hyperon worldline configurations. This gives
\begin{equation}
Z_{jk}^{}(N_t^{}) \equiv \sum_{\vec n_0^{}, \ldots, \vec n_{N_t}^{}}
\langle \psi_j^{} | \hat{\slashed M}^{N_t^{}}_{\{\vec n_j\}} | \psi_k^{} \rangle,
\end{equation}
where the product
\begin{equation}
\hat{\slashed M}^{N_t^{}}_{\{\vec n_j\}} \equiv
\hat{\slashed M}_{\vec n_{N_t}^{}, \vec n_{N_t-1}^{}}^{}
\hat{\slashed M}_{\vec n_{N_t-1}^{}, \vec n_{N_t-2}^{}}^{}
\ldots
\hat{\slashed M}_{\vec n_2^{}, \vec n_1^{}}^{}
\hat{\slashed M}_{\vec n_1^{}, \vec n_0^{}}^{},
\end{equation}
is expressed in terms of the reduced transfer matrices~(\ref{Mslash_stat}) and~(\ref{Mslash_hop}).
Here, $\vec n_j$ denotes the spatial position of the hyperon impurity (which has been integrated out)
on time slice $j$. The expressions for the projection amplitude and determinant are generalized to
\begin{equation}
Z_{jk}^{}(N_t^{}) = \sum_{\vec n_0^{}, \ldots, \vec n_{N_t}^{}} \mathrm{det} \, M_{p\times p}^{jk},
\end{equation}
where
\begin{equation}
M_{p\times p}^{jk} = \left(
\begin{array}{c c c} 
\vspace{.1cm}
\langle \phi_{0,j}^{} | \hat{\slashed M}^{N_t^{}}_{\{\vec n_j\}} | \phi_{0,k}^{} \rangle & 
\langle \phi_{0,j}^{} | \hat{\slashed M}^{N_t^{}}_{\{\vec n_j\}} | \phi_{1,k}^{} \rangle & \cdots \\
\vspace{.1cm}
\langle \phi_{1,j}^{} | \hat{\slashed M}^{N_t^{}}_{\{\vec n_j\}} | \phi_{0,k}^{} \rangle &
\langle \phi_{1,j}^{} | \hat{\slashed M}^{N_t^{}}_{\{\vec n_j\}} | \phi_{1,k}^{} \rangle & \cdots \\
\vdots & \vdots & \ddots
\end{array}
\right),
\end{equation}
such that the determinant is now to be computed over all possible hyperon wordline configurations.

We note that the worldline configuration is to be updated stochastically using a Metropolis algorithm. Thus, 
proposed changes in the impurity worldline are accepted or rejected by importance sampling
with $|Z_{jj}(N_t)|$ as the probability weight function. Here, $j$ denotes one of the initial trial nucleon cluster states.

\subsection{Worldline updates} 

\begin{figure}[h]
\begin{center}
\includegraphics[width = .85\columnwidth]{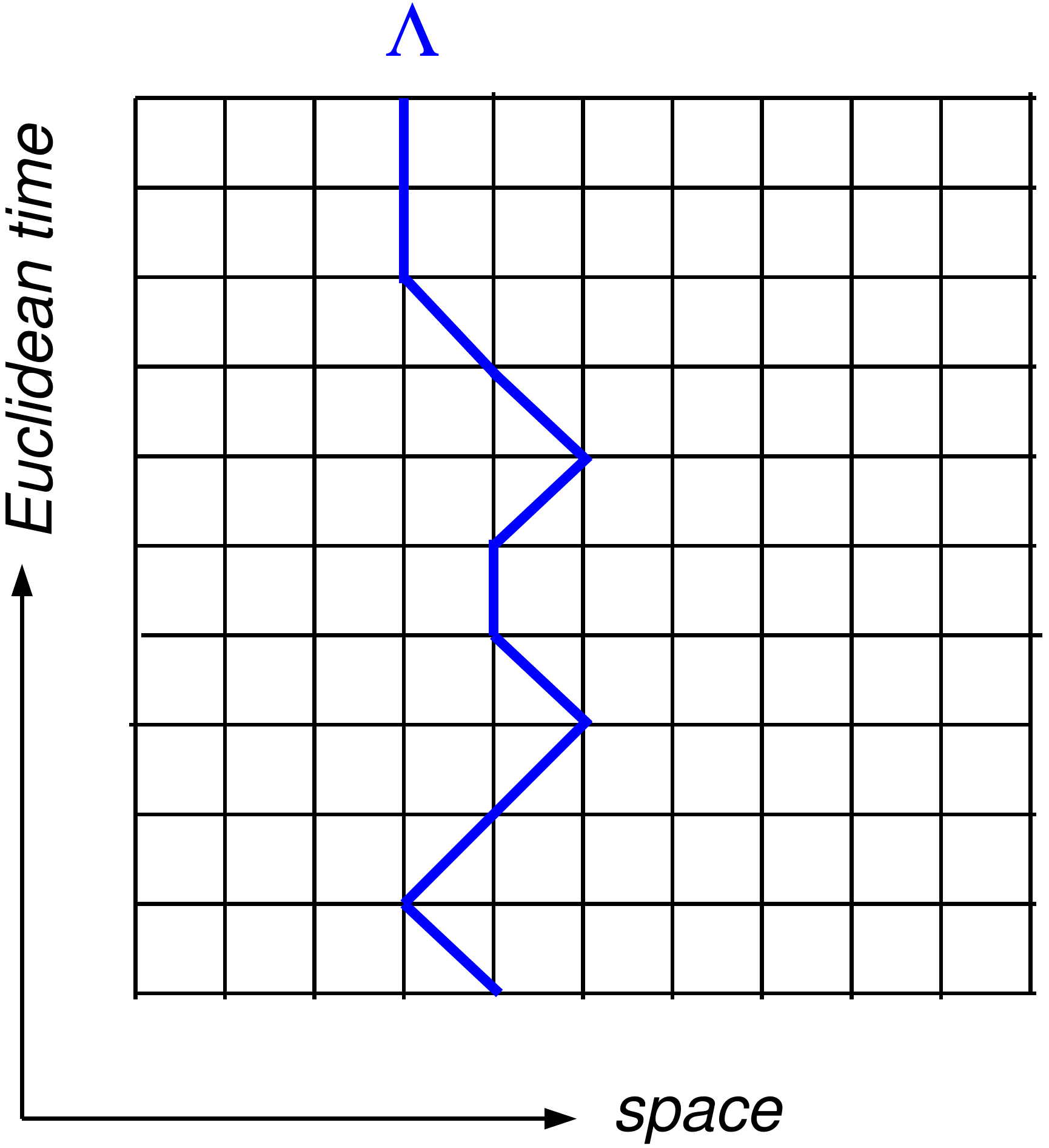}
\caption{Illustration of the hyperon worldline. In the reduced transfer matrix formalism, the hyperon has been ``integrated out'', and the interaction
between the hyperon and the nucleons is mediated by an effective ``background field'' generated by the hyperon worldline.
\label{fig_worldline}}
\end{center}
\end{figure}

The updating of the impurity worldline is handled in two steps: The generation of a new proposed worldline, and a Metropolis accept/reject 
step to determine whether to use the generated worldline. For this work, the worldline $W(\vec n, n_t^{})$ is a function of only the lattice 
site $\vec n$ and the Euclidean time step $n_t^{}$, and is equal to 1 where the impurity is present, and 0 at all other lattice points. 
From the expressions of the reduced transfer matrices, 
the worldline at two adjacent time steps, $W(\vec n^\prime,n_t^{})$ and $W^\prime(\vec n^\prime, n_t^{}+1)$ 
must obey the relation $|\vec n - \vec n^\prime | \leq 1$. For an illustration of the impurity (hyperon) worldline, see Fig.~\ref{fig_worldline}.

For the non-interacting worldline, we can generate new configurations from the free probabilities, as determined from the reduced 
transfer matrices. In this case, $P_h = h$ is the hopping probability, and $P_s = (1 - 6h)$ is the probability to remain stationary.
When initializing the worldline at the beginning of the MC simulation, we may start from a configuration 
where the worldline is completely stationary (``cold start'') or one where the worldline either hops or remains stationary at each time step according to the probabilities $P_h$ and $P_s$ (``warm start'').

At the beginning of every sweep through the lattice, we propose a new worldline to use for that sweep. This is done 
by taking the previous worldline and choosing a random time at which we cut the worldline and regenerating it either in the forwards and backwards time direction.  The new worldline is then accepted or rejected using a Metropolis accept or reject condition to preserve detailed balance associated with the absolute value of the amplitude.


\section{Results \label{sec:res}}

For the results presented in what follows, we use a spatial lattice spacing $a = 1/$(100~MeV) and temporal
lattice spacing of $a_t = 1/$(300~MeV).  The non-local smearing parameter is chosen to be $s_{\rm NL} = 0.2$,
and the local smearing parameter is set to $s_{\rm L} = 0.0$.  Since we only consider $s$-shell nuclei and
hypernuclei in this study, the local attraction provided by  $s_{\rm L}$ for heavier
nuclei is not needed~\cite{Elhatisari:2016owd}.  The coupling constant $C_{NN}$ is set to
$-7.5\times10^{-6}$~MeV$^{-2}$, and this combination of parameters yields a nucleon-nucleon scattering
length $a_{NN} = 6.86$~fm and effective range $r_{NN} = 1.77$~fm. 
The scattering length and effective range are calculated using Lüscher's finite volume method \cite{Luscher:1990ux}, as described in the Appendix of Ref.~\cite{Lee:2007ae}.  We find that these parameters produce good results for the average $S$-wave phase shifts as well as the three- and four-nucleon binding energies.  The exact transfer matrix calculation of the three-nucleon system and the Monte Carlo calculation of the four-nucleon system are both described in the following paragraphs. As stated previously, in this study the spin-dependent terms of the nucleon-nucleon interaction are not accounted for.  

For the $YN$ interaction, we set $C_{YN}$ according to the best overall fit to the light hypernuclei. Fitting to the $\Lambda$ separation energies for $^3_\Lambda$H, $^4_\Lambda$H/He, and $^5_\Lambda$He, we find $C_{YN} = -1.6 \times10^{-5}$~MeV$^{-2}$.  This gives $a_{YN} = -0.45$~fm for the scattering length and
$r_{YN} = -0.45$~fm for the effective range. In Table~\ref{hypertriton_benchmark}, we present benchmark
calculations of the ILMC results for $^3_\Lambda$H in comparison with exact
transfer matrix calculations. We show the results for the
energy as a function of Euclidean projection time.

\begin{table}[h]
\begin{center}
\caption{ILMC results for the energy of $^3_\Lambda$H versus Euclidean time in
    comparison with exact transfer matrix results for periodic box length 15.8~fm.
\label{hypertriton_benchmark}}
\bgroup
\def\arraystretch{1.2}
\begin{tabular}{|c|c|c|c|}
\hline
$N_t^{}$ & $t$ (MeV$^{-1}$) & ILMC (MeV) & Exact (MeV) \\
\hline
50 & 0.1667 & $-$1.0878(6) & $-$1.0878 \\
100 & 0.3333 & $-$1.4598(9) & $-$1.4590 \\
150 & 0.5000 &$-$1.6778(11) & $-$1.6760 \\
200 & 0.6667 &$-$1.7975(13) & $-$1.7966 \\
250 & 0.8333 &$-$1.8630(17) & $-$1.8614 \\
300 & 1.0000 & $-$1.8971(18) & $-$1.8954 \\
\hline
\end{tabular}
\egroup
\end{center}
\end{table}

We see that the agreement is quite good. The initial/final nucleon trial states for these calculations are taken to be spatially constant functions, which correspond to single-particle states of zero momentum in a periodic cubic box.  The hyperon initial/final
wave function is also taken be a constant function.  Since we use a constant initial/final state wave function for the hyperon, the initial/final positions for the hyperon worldline are irrelevant in the Monte Carlo updating process.  These exact transfer matrix calculations include
the induced three-baryon interaction described in Eq.~(\ref{H_spurious}).  

In Table~\ref{hypertriton_results}, we present exact Lanczos transfer matrix calculations of the
ground state of $^2$H, $^3_\Lambda$H, and separation energy $B_\Lambda^{}$, as a function of periodic box length.
In this work, we also present the exact Lanczos transfer matrix calculation wherever it is computationally
possible and using Monte Carlo for cases where it is not.  Given the extremely small $\Lambda$ separation energy,
it is necessary to go to very large volumes in order to remove finite volume artifacts.
Interestingly, $B_\Lambda^{}$ is found to be relatively constant with the periodic box size $L$.
This suppression of the finite volume dependence is an indication that the asymptotic normalization
coefficient of the hypertriton wave function is small~\cite{Konig:2011nz,Konig:2017krd}.
 
\begin{table}[h]
\begin{center}
\caption{Exact transfer matrix results for $^2$H, $^3_\Lambda$H, and the separation energy $B_\Lambda^{}$ versus periodic box length.
\label{hypertriton_results}}
\bgroup
\def\arraystretch{1.2}
\begin{tabular}{|c|c|c|c|}
\hline
$L$ (fm) & $^2$H (MeV) &  $^3_\Lambda$H (MeV)& $B_\Lambda$ (MeV)\\
\hline
15.8 & $-$1.651 & $-$1.932 & 0.281\\
17.8 & $-$1.460 & $-$1.712 & 0.252\\
19.7 & $-$1.332 & $-$1.569 & 0.237\\
21.7 & $-$1.245 & $-$1.474 & 0.228\\
23.7 & $-$1.186 & $-$1.410 & 0.224\\
25.6 & $-$1.146 & $-$1.368 & 0.222\\
27.6 & $-$1.118 & $-$1.339 & 0.221\\
29.6 & $-$1.100 & $-$1.319 & 0.220\\
\hline
\end{tabular}
\egroup
\end{center}
\end{table}

\begin{figure}[t]
        \begin{center}
        \includegraphics[width=6.17cm,angle=-90]{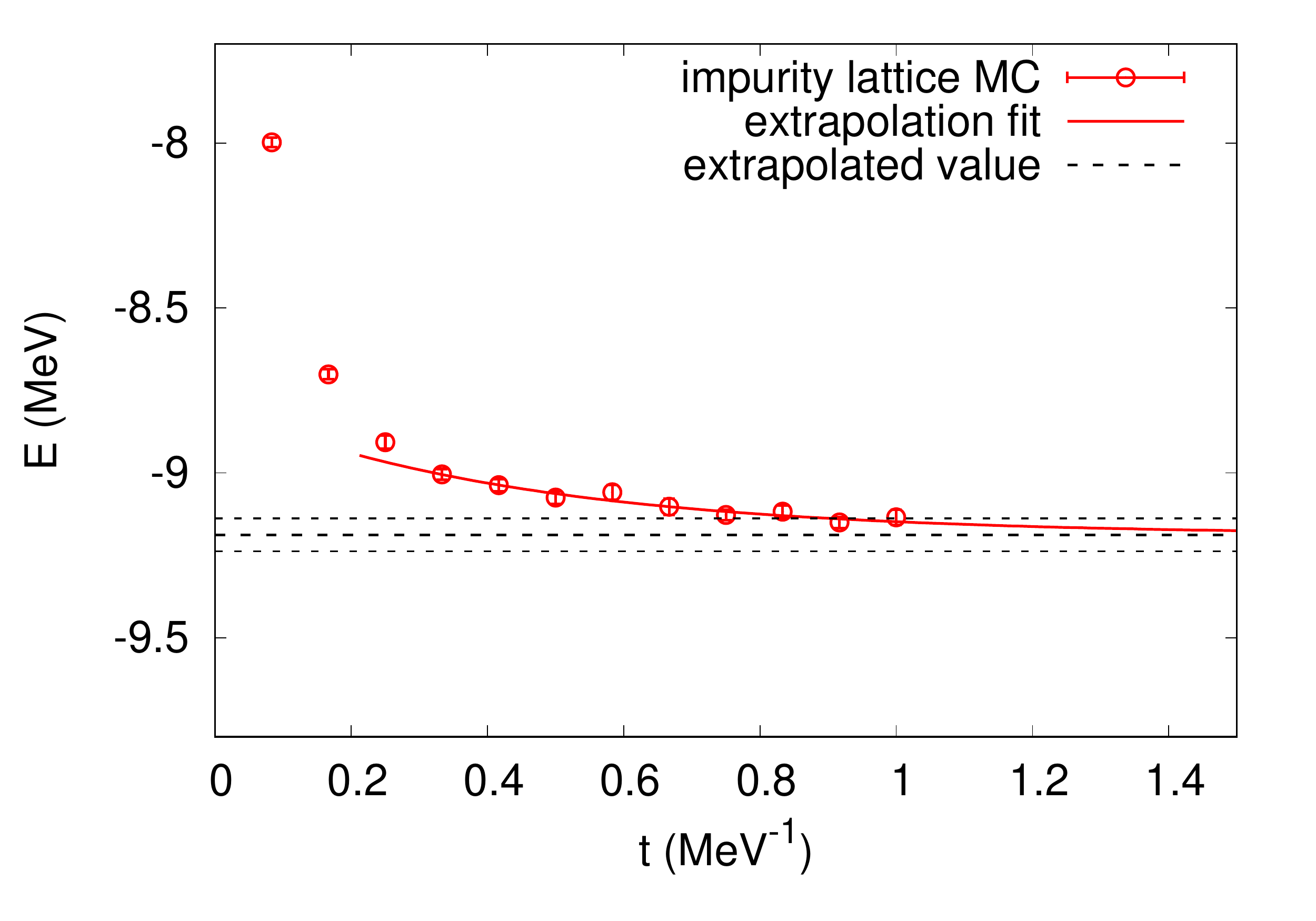}
        \caption{ILMC results for the $^4_{\Lambda}$H/He energy versus
          Euclidean projection time in a periodic box size of $L = 15.8$~fm.  We extract the ground state energy using
          an exponential \textit{ansatz} for the asymptotic time dependence.\label{4H_Lambda}}
        \end{center}
\end{figure}

In Fig.~\ref{4H_Lambda}, we present ILMC results for the $^4_{\Lambda}$H/He
energy versus Euclidean time. These calculations use a periodic box size of $L = 15.8$~fm with up to
$N_t = 300$ Euclidean time steps.  In order to extract the ground state energy, we use the extrapolation \textit{ansatz}
\begin{equation}
    E(t) = E_0^{} + c \exp(- \Delta E t),
    \label{exponential}
\end{equation}
which takes into account the residual dependence of the first excited state that couples to our
initial/final states. For this calculation, we use an initial/final state where the nucleon states have a spatially decaying exponential form with respect to the nucleus center of mass, while the initial/final hyperon wave function is a constant function.  It suffices to have an initial/final state with some overlap with the ground state wave function, and we find that these choices work very well. 

\begin{figure}[t]
        \begin{center}
        \includegraphics[width=6.17cm,angle=-90]{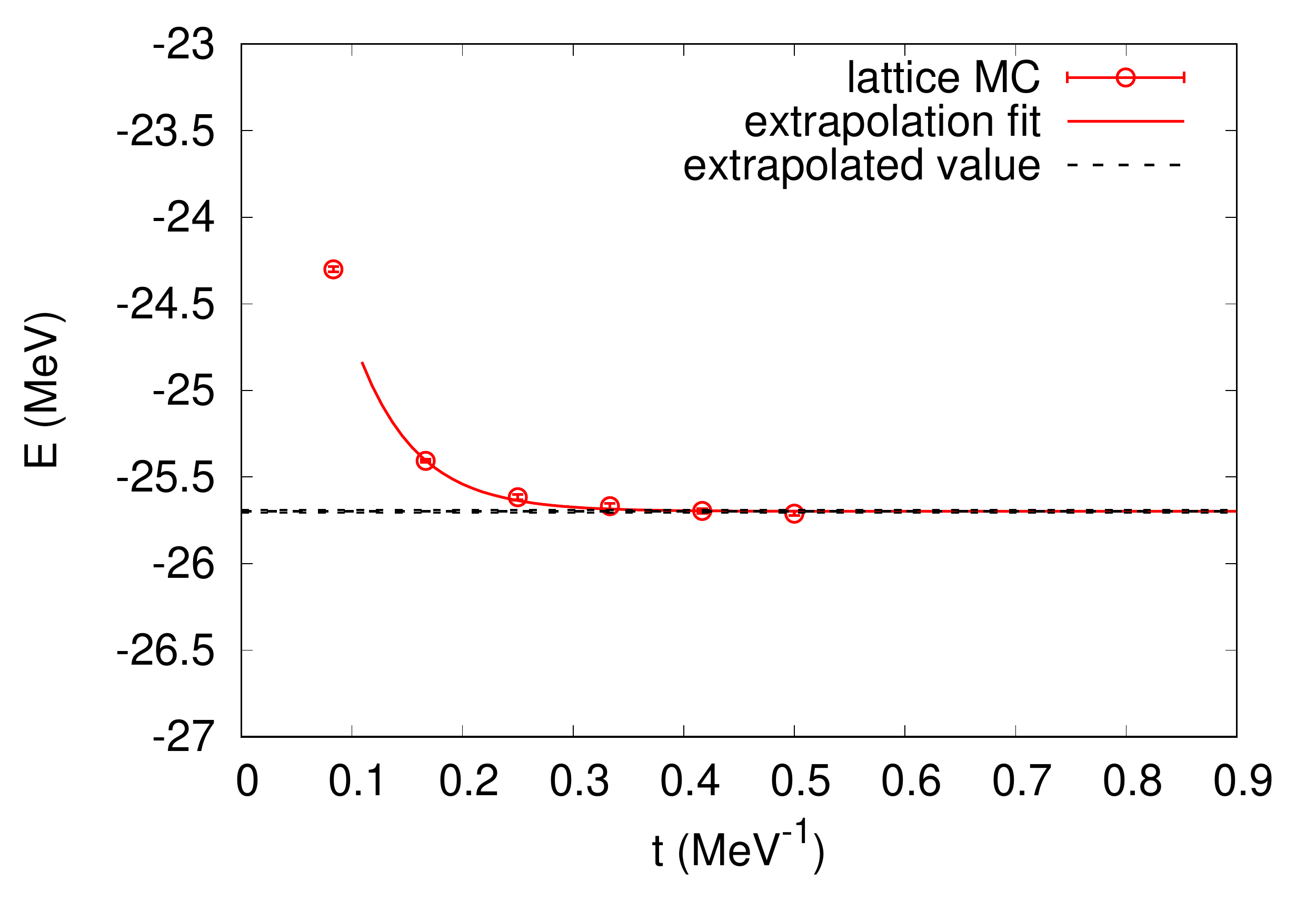}
        \caption{LMC results for the $^4$He energy versus Euclidean projection time in a periodic
          box size of $L = 9.9$~fm.  We extract the ground state energy using an exponential \textit{ansatz} for the
          asymptotic time dependence.   \label{4He}}
        \end{center}
\end{figure}

In Fig.~\ref{4He}, we show lattice Monte Carlo (LMC) results for the $^4$He energy versus Euclidean time.  As
there are no hyperons in this system, these are auxiliary field Monte Carlo calculations without impurity worldlines.  These calculations use a periodic box size of $L = 9.9$~fm with up to $N_t = 150$ Euclidean time steps. In order to extract the ground state energy,
we again use the exponential \textit{ansatz} in Eq.~(\ref{exponential}).  For this calculation, we again use an
initial/final state where the nucleons have a spatially-decaying exponential form with respect to the nucleus
center of mass.

\begin{figure}[t]
        \begin{center}
        \includegraphics[width=6.17cm,angle=-90]{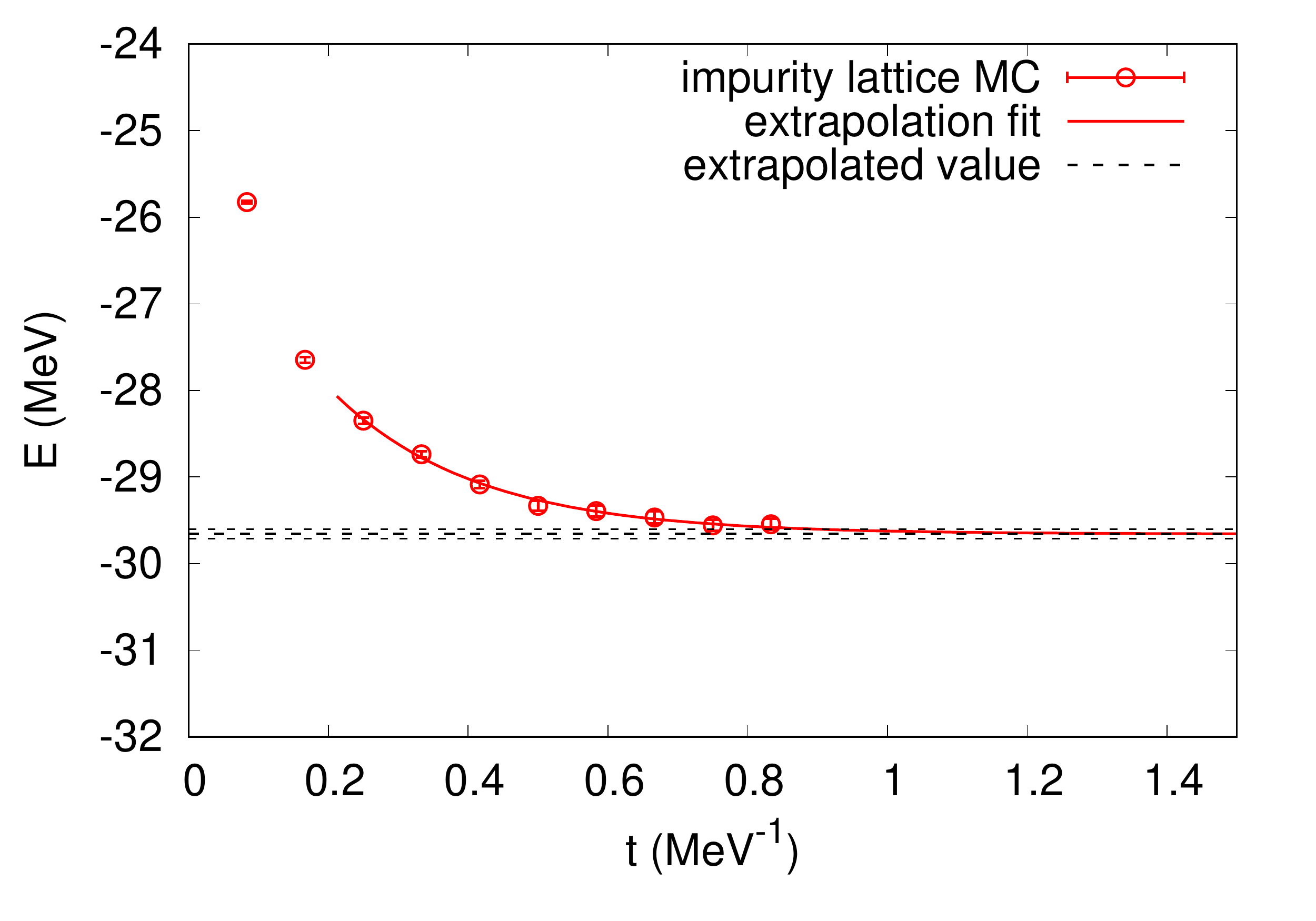}
        \caption{ILMC results for the $^5_\Lambda$He energy versus Euclidean time
          in a periodic box size of $L = 9.9$~fm.  We extract the ground state energy using an exponential
          \textit{ansatz} for the asymptotic time dependence. \label{5He_Lambda}}
        \end{center}
\end{figure}

In Fig.~\ref{5He_Lambda}, ILMC results are shown for the $^5_\Lambda$He energy versus
Euclidean time.  These calculations use a periodic box size of $L = 9.9$~fm with up to $N_t = 250$ Euclidean time
steps.  We again use the exponential \textit{ansatz} from Eq.~(\ref{exponential}) to extract the ground state energy.
Similar to the $^4_{\Lambda}$H/He calculation, here we use an initial/final state where the nucleons have a
spatially decaying exponential form with respect to the nucleus center of mass, while the initial/final hyperon
wave function is a constant function. 

In Table~\ref{collected_results}, we present the lattice results for all of the $s$-shell nuclei and
hypernuclei.  The exact transfer matrix results are shown without error bars, while the ILMC 
and LMC results are shown with error bars that take into account
stochastic errors and extrapolation errors.  There is also a residual systematic error due to finite
volume effects.  For a box size of $L = 29.6$~fm, the finite volume error on $^2$H is $0.04$~MeV, and the
estimated finite volume error for $^3_\Lambda$H is also $\simeq 0.04$~MeV.  As both corrections are in the same
direction (with more binding at finite volume), the resulting finite volume error on the separation energy
is $<0.002$~MeV.

For a box size of $L = 15.8$~fm, the finite volume error on $^3$H/He is $\simeq 0.10$~MeV, and the estimated finite
volume errors for $^4_\Lambda$H/He are also $\simeq 0.10$~MeV. For a box size of $L = 9.9$~fm, the finite volume
error on $^4$He is $\simeq 1.5$~MeV, and the estimated finite volume errors for $^4_\Lambda$H/He are $\simeq 2.0$~MeV.

\begin{table}[h]
\begin{center}
  \caption{Summary of lattice results (exact transfer matrix, ILMC and LMC) for the energies of light nuclei and hypernuclei, and for separation energies. Comparisons with experimental separation energies are given where such data exists. These comparisons are averaged over 
    Wigner SU(4) and $\Lambda$ spin components. For the case of $^4_\Lambda$H/He, we average over the $0^+$ and $1^+$ separation energies for $^4_\Lambda$H and $^4_\Lambda$He weighted by number of spin components.  More data can be found in the review Ref.~\cite{Davis:2005mb}.
\label{collected_results}}
\bgroup
\begin{adjustbox}{width=8.5cm}
\def\arraystretch{1.2}
\begin{tabular}{|c|c|c|c|c|}
\hline
Nucleus & $L$ (fm)  & $E$ (MeV) & $B_\Lambda$ (MeV) & $B^{\rm exp}_\Lambda$ (MeV)\\
\hline
$^2$H & 29.6 & $-$1.100 & -- & --\\
$^3_\Lambda$H & 29.6 & $-$1.319 & 0.220 & 0.13(5) \cite{Bohm:1968qkc,Juric:1973zq,Davis:1991zpu}\\
$^3$H/He & 15.8 & $-$8.725 & -- & -- \\
$^4_\Lambda$H/He & 15.8 & $-$9.19(5) & 0.46(5) & 1.39(4) \cite{Bohm:1968qkc,Juric:1973zq,Davis:1991zpu,Bamberger:1973ht,Bedjidian:1979qh}\\
$^4$He & 9.9 & $-$25.698(9) & -- & --\\
$^5_\Lambda$He & 9.9 & $-$29.66(6) & 3.96(6)& 3.12(2) \cite{Bohm:1968qkc,Juric:1973zq,Davis:1991zpu} \\
\hline
\end{tabular}
\end{adjustbox}
\egroup
\end{center}
\end{table}

For the comparison with the experimental results, we average over Wigner SU(4) and $\Lambda$ spin components where data exists.  We see that while the $B^{\rm exp}_\Lambda$ is larger than the experimental values for
$^3_\Lambda$H and $^5_\Lambda$He, the separation is smaller than experimental value for $^4_\Lambda$H/He.
This is an indication that there are deficiencies in our very simple treatment of the $YN$
and $NN$ interactions.  However, this serves as a good starting point for determining the
essential features of the $YN$ interactions needed to describe the structure and properties
of hypernuclei.


\section{Discussion \label{sec:disc}}

We have shown, as a proof of principle, how state-of-the-art NLEFT calculations can be extended
to include hyperons. As the number of hyperons in realistic hypernuclei is small (typically one or two)
relative to the number of nucleons, we have applied the ILMC method whereby the hyperon
``impurity'' is integrated out and represented by a hyperon ``worldline'', the position of which
is updated during the MC calculation. Effectively, the standard NLEFT calculations for nucleons
are augmented by a ``background field'' induced by the hyperon worldline. We have benchmarked
the ILMC method by presenting preliminary MC 
results for the $s$-shell hypernuclei, using a simplified interaction similar to pionless EFT.

One of the most promising aspects of this work is the fact that the ILMC simulations scale very favorably with the number of nucleons. We have found that nearly all of
the computational effort is consumed in calculating single-nucleon amplitudes as a function of
the auxiliary field.  As this part of the code scales linearly with the number of nucleons,
it should be possible to perform calculations of hypernuclei with up to one hundred or more nucleons.
We note also that the particular set of interactions that we have used here can also be directly
applied to studying the properties of a bosonic impurity immersed in a superfluid Fermi gas.
By modifying the included $P$-wave interactions of the impurity, we would also be able to describe
the properties of an alpha particle immersed in a gas of superfluid neutrons.  The possible applications
of this method clearly go well beyond hypernuclear structure calculations and have general utility
for numerous quantum many-body systems.

Returning to hypernuclear systems, the obvious next extension of this work is to
include spin-dependent $YN$ interactions.  The importance of the spin-dependence of the $YN$
interaction can be seen clearly in the splittings between the $0^+$ and $1^+$ states in $^4_\Lambda$H and
$^4_\Lambda$He Ref.~\cite{Gibson:1995an}.  One should also include explicit $\Lambda N$-$\Sigma N$ transitions,
see \textit{e.g}.~\cite{Beane:2003yx}, as well as one-meson exchange interactions that would put the
$YN$ interaction in the same EFT formalism \cite{Haidenbauer:2013oca,Haidenbauer:2019boi} as currently
used for the $NN$ interaction in NLEFT~\cite{Li:2018ymw}. 

The number of adjustable parameters in the $YN$ interaction will then increase. The most natural
approach, in line with the treatment of the $NN$ interaction, would be to fit such parameters 
to $\Lambda N$ scattering phase shifts. However, due to the paucity of such data (especially at
low energies), we expect to need at least the hypertriton binding 
energy as an additional constraint, as it is also done in continuum chiral EFT, see \textit{e.g.} 
Ref.~\cite{Haidenbauer:2019boi}.  As the effects of $\Lambda N$-$\Sigma N$ transitions are included,
it may be necessary to use further empirical data on other light hypernuclei to constrain the relevant LECs. A further extension
concerns the extension to $S=-2$ hypernuclei, which on the one hand would
involve the $YY$ interactions~\cite{Polinder:2007mp,Haidenbauer:2015zqb,Hiyama:2018lgs}
and on the other hand a  modified ILMC algorithm for two interacting worldlines.
Work along these lines is underway.


\section*{Acknowledgments}

We thank Avraham Gal, Hoai Le, Ning Li, Bing-Nan Lu and Andreas Nogga for useful discussions.
This work was supported by DFG and NSFC through funds provided to the
Sino-German CRC 110 ``Symmetries and the Emergence of Structure in QCD" (NSFC
Grant No.~11621131001, DFG Grant No.~TRR110).
The work of UGM was supported in part by VolkswagenStiftung (Grant no. 93562)
and by the CAS President's International
Fellowship Initiative (PIFI) (Grant No.~2018DM0034).
The work of DL is supported in part by the U.S. Department of Energy (Grant 
No. DE-SC0018638) and the Nuclear Computational Low-Energy Initiative (NUCLEI) SciDAC project.
The authors gratefully acknowledge the Gauss Centre for Supercomputing e.V. (www.gauss-centre.eu) 
for funding this project by providing computing time on the GCS Supercomputer JUWELS 
at the J\"ulich Supercomputing Centre (JSC).



\begin{thebibliography}{99}

\bibitem{Korpa:2001au}
C.~Korpa, A.~Dieperink and R.~Timmermans,
Phys. Rev. C \textbf{65} (2002) 015208.
  
\bibitem{Polinder:2006zh}
H.~Polinder, J.~Haidenbauer and U.-G.~Mei{\ss}ner,
Nucl. Phys. A \textbf{779} (2006) 244.

\bibitem{Lee:2008fa}
D.~Lee,
Prog. Part. Nucl. Phys. \textbf{63} (2009) 117.

\bibitem{Lahde:2019npb}
T.~A.~L\"ahde and U.-G.~Mei{\ss}ner,
Lect. Notes Phys. \textbf{957} (2019), 1.




\bibitem{Haidenbauer:2013oca}
J.~Haidenbauer, S.~Petschauer, N.~Kaiser, U.-G.~Mei{\ss}ner, A.~Nogga and W.~Weise,
  Nucl.\ Phys.\ A {\bf 915} (2013) 24.

\bibitem{Haidenbauer:2019boi}
J.~Haidenbauer, U.-G.~Mei{\ss}ner and A.~Nogga,
Eur. Phys. J. A \textbf{56} (2020) 91.

\bibitem{Lonardoni:2013rm}
D.~Lonardoni, S.~Gandolfi and F.~Pederiva,
Phys. Rev. C \textbf{87} (2013) 041303.

\bibitem{Gazda:2016qva}
D.~Gazda and A.~Gal,
Nucl. Phys. A \textbf{954} (2016) 161.

\bibitem{Wirth:2017lso}
R.~Wirth and R.~Roth,
Phys. Lett. B \textbf{779} (2018) 336.

\bibitem{Wirth:2017bpw}
R.~Wirth, D.~Gazda, P.~Navr\'atil and R.~Roth,
Phys. Rev. C \textbf{97} (2018)  064315.

\bibitem{Le:2019gjp}
H.~Le, J.~Haidenbauer, U.-G.~Mei{\ss}ner and A.~Nogga,
Phys. Lett. B \textbf{801} (2020) 135189.


\bibitem{Haidenbauer:2019thx}
J.~Haidenbauer and I.~Vidana,
Eur. Phys. J. A \textbf{56} (2020) 55.




\bibitem{Bour}S.~Bour, MSc thesis, University of Bonn (2009).

\bibitem{Wagman:2017tmp}
M.~L.~Wagman, F.~Winter, E.~Chang, Z.~Davoudi, W.~Detmold, K.~Orginos, M.~J.~Savage and P.~E.~Shanahan,
Phys. Rev. D \textbf{96} (2017) 114510 


\bibitem{Elhatisari:2014lka}
S.~Elhatisari and D.~Lee,
  Phys.\ Rev.\ C {\bf 90} (2014)  064001.

\bibitem{Bour:2014bxa}
S.~Bour, D.~Lee, H.~W.~Hammer and U.-G.~Mei{\ss}ner,
Phys. Rev. Lett. \textbf{115} (2015)  185301

\bibitem{Lu:2018bat}
B.~N.~Lu, N.~Li, S.~Elhatisari, D.~Lee, E.~Epelbaum and U.-G.~Mei{\ss}ner,
Phys. Lett. B \textbf{797} (2019) 134863.

\bibitem{Hammer:2001ng}
H.~Hammer,
Nucl. Phys. A \textbf{705} (2002) 173.

\bibitem{Contessi:2018qnz}
L.~Contessi, N.~Barnea and A.~Gal,
Phys. Rev. Lett. \textbf{121} (2018)  102502.


\bibitem{Elhatisari:2016owd}
S.~Elhatisari {\it et al.},
Phys. Rev. Lett. \textbf{117} (2016)  132501.



\bibitem{Elhatisari:2017eno}
S.~Elhatisari {\it et al.},
Phys. Rev. Lett. \textbf{119} (2017)  222505.

\bibitem{Li:2018ymw}
N.~Li, S.~Elhatisari, E.~Epelbaum, D.~Lee, B.~N.~Lu and U.-G.~Mei{\ss}ner,
Phys. Rev. C \textbf{98} (2018)  044002.

\bibitem{Wigner:1936dx}
E.~Wigner,
Phys. Rev. \textbf{51} (1937) 106.

\bibitem{Luscher:1990ux}
M.~L\"uscher,
Nucl. Phys. B \textbf{354} (1991) 531.

\bibitem{Lee:2007ae}
D.~Lee,
Eur. Phys. J. A \textbf{35} (2008) 171.


\bibitem{Konig:2011nz}
S.~K\"onig, D.~Lee and H.~W.~Hammer,
Phys. Rev. Lett. \textbf{107} (2011) 112001.

\bibitem{Konig:2017krd}
S.~K\"onig and D.~Lee,
Phys. Lett. B \textbf{779} (2018) 9.

\bibitem{Gibson:1995an}
B.~F.~Gibson and E.~V.~Hungerford,
Phys. Rept. \textbf{257} (1995) 349.

\bibitem{Bohm:1968qkc}
G.~Bohm,  {\it et al.},
Nucl. Phys. B \textbf{4} (1968) 511.

\bibitem{Juric:1973zq}
M.~Juric,  {\it et al.},
Nucl. Phys. B \textbf{52} (1973) 1.

\bibitem{Davis:1991zpu}
D.~H.~Davis,
AIP Conf. Proc. \textbf{224} (1991) 38.

\bibitem{Bamberger:1973ht}
A.~Bamberger \textit{et al.} [CERN-Heidelberg-Warsaw],
Nucl. Phys. B \textbf{60} (1973) 1.

\bibitem{Bedjidian:1979qh}
M.~Bedjidian \textit{et al.} [CERN-Lyon-Warsaw],
Phys. Lett. B \textbf{83} (1979) 252.

\bibitem{Davis:2005mb}
D.~H.~Davis,
Nucl. Phys. A \textbf{754} (2005) 3.

\bibitem{Beane:2003yx}
S.~Beane, P.~Bedaque, A.~Parreno and M.~Savage,
Nucl. Phys. A \textbf{747} (2005) 55.


\bibitem{Polinder:2007mp}
H.~Polinder, J.~Haidenbauer and U.-G.~Mei{\ss}ner,
Phys. Lett. B \textbf{653} (2007) 29.


\bibitem{Haidenbauer:2015zqb}
J.~Haidenbauer, U.-G.~Mei{\ss}ner and S.~Petschauer,
Nucl. Phys. A \textbf{954} (2016) 273.

\bibitem{Hiyama:2018lgs}
E.~Hiyama and K.~Nakazawa,
Ann. Rev. Nucl. Part. Sci. \textbf{68} (2018) 131.

\end{thebibliography}
\end{document}